\documentstyle[11pt,epsfig]{article}
\newcounter{subequation}[equation]

\makeatletter 
 
\def\thesubequation{\theequation\@alph\c@subequation}
\def\@subeqnnum{{\rm (\thesubequation)}}
\def\slabel#1{\@bsphack\if@filesw {\let\thepage\relax
   \xdef\@gtempa{\write\@auxout{\string
      \newlabel{#1}{{\thesubequation}{\thepage}}}}}\@gtempa
   \if@nobreak \ifvmode\nobreak\fi\fi\fi\@esphack}
\def\subeqnarray{\stepcounter{equation}
\let\@currentlabel=\theequation\global\c@subequation\@ne
\global\@eqnswtrue
\global\@eqcnt\z@\tabskip\@centering\let\\=\@subeqncr
$$\halign to \displaywidth\bgroup\@eqnsel\hskip\@centering
  $\displaystyle\tabskip\z@{##}$&\global\@eqcnt\@ne
  \hskip 2\arraycolsep \hfil${##}$\hfil
  &\global\@eqcnt\tw@ \hskip 2\arraycolsep
  $\displaystyle\tabskip\z@{##}$\hfil
   \tabskip\@centering&\llap{##}\tabskip\z@\cr}
\def\endsubeqnarray{\@@subeqncr\egroup
                     $$\global\@ignoretrue}
\def\@subeqncr{{\ifnum0=`}\fi\@ifstar{\global\@eqpen\@M
    \@ysubeqncr}{\global\@eqpen\interdisplaylinepenalty \@ysubeqncr}}
\def\@ysubeqncr{\@ifnextchar [{\@xsubeqncr}{\@xsubeqncr[\z@]}}
\def\@xsubeqncr[#1]{\ifnum0=`{\fi}\@@subeqncr
   \noalign{\penalty\@eqpen\vskip\jot\vskip #1\relax}}
\def\@@subeqncr{\let\@tempa\relax
    \ifcase\@eqcnt \def\@tempa{& & &}\or \def\@tempa{& &}
      \else \def\@tempa{&}\fi
     \@tempa \if@eqnsw\@subeqnnum\refstepcounter{subequation}\fi
     \global\@eqnswtrue\global\@eqcnt\z@\cr}
\let\@ssubeqncr=\@subeqncr
\@namedef{subeqnarray*}{\def\@subeqncr{\nonumber\@ssubeqncr}\subeqnarray}
\@namedef{endsubeqnarray*}{\global\advance\c@equation\m@ne%
                           \nonumber\endsubeqnarray}

\def\dalemb#1#2{{\vbox{\hrule height .#2pt
        \hbox{\vrule width.#2pt height#1pt \kern#1pt
                \vrule width.#2pt}
        \hrule height.#2pt}}}
\def\square{\mathord{\dalemb{6.8}{7}\hbox{\hskip1pt}}}

    \let\e=\epsilon
  \let\q=\theta  
  \let\n=\nu

\def\nn{\nonumber} \def\bd{\begin{document}} \def\ed{\end{document}}
\def\ds{\documentstyle} \let\fr=\frac \let\bl=\bigl \let\br=\bigr
\let\Br=\Bigr \let\Bl=\Bigl 
\let\bm=\bibitem
\let\na=\nabla
\let\pa=\partial \let\ov=\overline
\def\ie{{\it i.e.\ }} 
\newcommand{\be}{\begin{equation}} 
\newcommand{\ee}{\end{equation}} 
\def\ba{\begin{array}}
\def\ea{\end{array}}
\def\ft#1#2{{\textstyle{{\scriptstyle #1}\over {\scriptstyle #2}}}}
\def\fft#1#2{{#1 \over #2}}
\def\del{\partial}
\def\sst#1{{\scriptscriptstyle #1}}
\def\oneone{\rlap 1\mkern4mu{\rm l}}
\def\e7{E_{7(+7)}}
\def\td{\tilde}
\def\wtd{\widetilde}
\def\im{{\rm i}}
\def\bog{Bogomol'nyi\ }
\def\q{{\tilde q}}
\def\hast{{\hat\ast}}
\def\0{{\sst{(0)}}}
\def\1{{\sst{(1)}}}
\def\2{{\sst{(2)}}}
\def\3{{\sst{(3)}}}
\def\4{{\sst{(4)}}}
\def\5{{\sst{(5)}}}
\def\6{{\sst{(6)}}}
\def\7{{\sst{(7)}}}
\def\8{{\sst{(8)}}}
\def\n{{\sst{(n)}}}
\def\oo{{\"o}}
\def\hA{\hat{\cal A}}
\def\ns{{\sst {\rm NS}}}
\def\rr{{\sst {\rm RR}}}
\def\tH{{\widetilde H}}
\def\tB{{\widetilde B}}
\def\cA{{\cal A}}
\def\cF{{\cal F}}
\def\tF{{\wtd F}}
\def\Z{\rlap{\sf Z}\mkern3mu{\sf Z}}
\def\ep{{\epsilon}}
\def\IIA{{\rm IIA}}
\def\IIB{{\rm IIB}}
\def\ads{{\rm AdS}}
\def\R{\rlap{\rm I}\mkern3mu{\rm R}}
\def\mapright#1{\smash{\mathop{-\!\!\!-\!\!\!-\!\!\!-\!\!\!-\!\!\!
             \longrightarrow}\limits^{#1}}}

\def\Ei{{\hbox{Ei}}}
\def\Ci{{\hbox{Ci}}}
\def\Si{{\hbox{Si}}}

\newcommand{\ho}[1]{$\, ^{#1}$}
\newcommand{\hoch}[1]{$\, ^{#1}$}
\newcommand{\bea}{\begin{eqnarray}} 
\newcommand{\eea}{\end{eqnarray}} 
\newcommand{\ra}{\rightarrow}
\newcommand{\lra}{\longrightarrow}
\newcommand{\Lra}{\Leftrightarrow}
\newcommand{\aap}{\alpha^\prime}
\newcommand{\bp}{\tilde \beta^\prime}
\newcommand{\tr}{{\rm tr} }
\newcommand{\Tr}{{\rm Tr} } 
\newcommand{\NP}{Nucl. Phys. }
\newcommand{\tamphys}{\it Center for Theoretical Physics,
Texas A\&M University, College Station, TX 77843}

\newcommand{\upenn}{\it Department of Physics and Astronomy,\\ University
of Pennsylvania, Philadelphia, PA 19104}

\newcommand{\brussels}{\it Physique Th\'eorique et Math\'ematique, 
Universit\'e Libre de Bruxelles,\\ Campus Plaine C.P. 231, B-1050
Bruxelles, Belgium} 

\newcommand{\auth}{H. L\"u\hoch{\dagger1} and J.F. 
V\'azquez-Poritz\hoch{\ddagger2}}

\begin{document}
\begin{flushright}
MCTP-02-05\\
ULB-TH/02-04\\
February  2002\\
\hfill{\bf hep-th/0202075}\\
\end{flushright}

\vspace{10pt}

\begin{center}

{\large {\bf Resolution of Overlapping Branes}}

\vspace{20pt}
\auth

\vspace{10pt}
{\hoch{\dagger}\it Michigan Center for Theoretical Physics\\
University of Michigan, Ann Arbor, Michigan 48109}

\vspace{10pt}
{\hoch{\ddagger}\brussels}

\vspace{30pt}

\underline{ABSTRACT}
\end{center}

     We obtain singularity resolutions for various overlapping brane
configurations, including those of two heterotic 5-branes, type II
5-branes or D4-branes.  In these solutions, the ``harmonic'' function
$H$ for each brane component depends only on the associated
four-dimensional relative transverse space.  The resolution is
achieved by replacing these transverse spaces with Eguchi-Hanson or
Taub-NUT spaces, both of which admit a normalisable self-dual (or
anti-self-dual) harmonic 2-form.  Due to the manner in which the
interaction terms for the form fields modify their Bianchi identities
or equations of motion, these normalisable harmonic 2-forms provide
regular sources for the branes.  We also obtain resolved 5-branes and
D4-branes wrapped on $S^1$, which is fibred over the transverse
Eguchi-Hanson or Taub-NUT spaces.  The T-duality invariance of the
NS-NS 5-brane is retained after the resolution.  The resolved 5-branes
and D4-branes provide regular supergravity duals of certain
supersymmetric Yang-Mills theories in five and four dimensions.

{\vfill\leftline{}\vfill
\vskip 10pt 
\footnoterule {\footnotesize \hoch{1} 
Research supported in full by DOE grant DE-FG02-95ER40899.

{\footnotesize \hoch{2} Research supported in full by the Francqui
Foundation (Belgium), the Actions de Recherche Concert{\'e}es 
\phantom{of the} of the
Direction de la Recherche Scientifique - Communaut\'e Francaise de
Belgique, IISN-Belgium
\phantom{of the} (convention 4.4505.86).  }
\vskip  -12pt}

\pagebreak
\setcounter{page}{1}


\section{Introduction}

     BPS branes play an important role in string and M-theory.
Certain supergravity solutions provide an explicit demonstration of
various properties in dual gauge theories.  However, in most cases
such a solution exhibits a singularity at the origin of the brane,
which imposes a severe restriction on the range of validity.  It is
expected that higher-order stringy terms or non-perturbative effects
can resolve these singularities, while maintaining a fixed mass/charge
ratio.

     A typical $p$-brane or intersecting $p$-brane solution makes use
of only the kinetic terms of the form fields, with zero contribution
from the interacting terms.  The interactions between form fields
modify their Bianchi identities and/or equations of motion, {\it
e.g.}
\be
dF_{\sst{(n)}}=F_{\sst{(p)}}\wedge F_{\sst{(q)}}\qquad\hbox{and/or}
\qquad d{\ast F_{\sst{(n)}}} = F_{\sst{(s)}}\wedge F_{\sst{(t)}}
\,.
\ee
In many cases, the inclusion of such a contribution can resolve brane
singularities at the level of supergravity.  An early example of this
is the heterotic 5-brane constructed in \cite{strom}, where the
5-brane source is the matter $SU(2)$ Yang-Mills instanton living in
the four-dimensional Euclidean transverse space.  This construction
makes use of the Bianchi identity $dF_\3 = \ft12 F_\2^i\wedge F_\2^i$
of the heterotic string theory.  The solution is expected to be valid
at all orders of string perturbation.  This construction can also be
extended to the case where the transverse space is K3, since one can
equate the $SU(2)$ Yang-Mills fields to the self-dual spin connection
of the K3 \cite{cd,strom,dmw}.  An alternative resolution makes use of
the well-known fact that the K3 manifold can be constructed from the
orbifold $T^4/Z_2$, with the sixteen fixed points each blown up using
the Eguchi-Hanson metric \cite{gp,page}.  The Eguchi-Hanson metric has
one normalisable self-dual 2-form, which can be set equal to a $U(1)$
2-form field strength associated with one of the sixteen Cartan
generators of the heterotic string.  The resulting 5-brane on the K3
is then regular everywhere \cite{clptrans}.

     Recently, a resolution procedure which makes use of interacting
terms in the Bianchi identity or equations of motion has been
extensively discussed for the D3-brane solutions of type IIB theory
\cite{klebtsey,klebstra,ganpol,gubser,zaytse,fre} and branes in type
IIA and M-theory
\cite{clptrans,cglprf,hk,cglpd2n2,cglphyper,cglpspin7,cgllp}.  These
non-singular examples may provide important supergravity dual
solutions of supersymmetric field theories on the worldvolume of the
branes, with the conformal symmetry typically broken by the
resolution.

      So far, the study of brane resolution has focused on a single
brane or, more precisely, on coincidental branes of the same type.  Branes
can also intersect with each other \cite{pt,tse,houart}.  For standard
intersections, the harmonic functions corresponding to each brane
depend on the overall transverse space. For non-standard
intersections, which are also called overlapping branes, the harmonic
functions depend only on the relative transverse spaces
\cite{bbj,gkt,edel}.

       In this letter, we find that, for the overlap of two heterotic
5-branes as well as that of two D4-branes, the two relative transverse
spaces are four-dimensional Ricci-flat spaces, and hence can be
replaced by the Eguchi-Hanson or Taub-NUT metrics.  Since the
Eguchi-Hanson and Taub-NUT metrics both support one normalisable
self-dual (or anti-self-dual, depending on the orientation) 2-form, we
can resolve these two overlapping branes by making use of the
corresponding Bianchi identities.  We obtain resolved overlapping
heterotic 5-brane and D4-brane solutions in sections 2 and 3
respectively.  In section 4, by performing the T-duality on the D4/D4
system, we obtain a resolved 5-brane overlap in type II theories. In
addition, we obtain regular 5-brane, D5-brane and D4-brane wrapped on an $S^1$,
which is fibred over the transverse Eguchi-Hanson or Taub-NUT spaces.
We conclude our paper in section 5.  In the appendix, we present
certain properties of Eguchi-Hanson and Taub-NUT metrics that are used
extensively throughout the letter.

        It should be emphasised that, although results are presented
here mostly using the Eguchi-Hanson metric, the construction also
works when it is replaced with the Taub-NUT metric of an appropriate
orientation.

\section{Overlapping heterotic 5-branes}

    The Lagrangian for the bosonic sector of ten-dimensional heterotic
supergravity is given by
\be
{\cal L}_{\rm het}=R{\ast \oneone}-
\ft12{\ast d\phi}\wedge d\phi-\ft12{\rm
e}^{-\phi}{\ast F_\3}\wedge F_\3-\ft12{\rm e}^{-\ft12\phi}
{\ast F_\2^i} \wedge F_\2^i\,,
\ee
where
\bea
F_\3&=&dA_\2+\ft12 A_\1^i \wedge
dA_\1^i+\ft16 f_{ijk}\,A_\1^i \wedge A_\1^j \wedge A_\1^k\,,\nn\\
F_\2^i&=&dA_\1^i +\ft12 f_{jk}^i\, A_\1^j \wedge A_\1^k\,.
\eea
Consider the solution describing a non-standard intersection of two
heterotic 5-branes \cite{khuri,bbj} for the case in which the
Yang-Mills fields $A_\1^i$ are set to zero:
\bea
ds_{10}^2&=&(H\, {\wtd H})^{-\ft14}(-dt^2 +dw^2+H\, dy_i^2+
{\wtd H}\, d{\td y}_i^2) \,,\nn\\
{\rm e}^{-\phi}{\ast F}_\3&=&{\wtd H}\, dt\wedge 
dw\wedge d^4{\td y}\wedge dH^{-1}+
H\, dt\wedge dw\wedge d^4y\wedge d{\wtd H}^{-1}\,,\nn\\
\phi&=&\ft12 {\rm log}(H\,{\wtd H})\,,\label{hethet}
\eea
where
\be
\square H=0\,,\qquad \wtd {\square} {\wtd H}=0\,,
\ee
and $\square$ ($\wtd {\square}$) is taken over the $y_i$ (${\td y}_i$)
directions.  The solution can be represented diagrammatically as
follows:

\bigskip\bigskip
\centerline{
\begin{tabular}{c|ccccccccccc}
&$t$ & $w$ & $y_1$ & $y_2$ & $y_3$ & $y_4$ & ${\tilde y}_1$ & 
${\tilde y}_2$ & ${\tilde y}_3$ & ${\tilde y}_4$ & \\ \hline
het 5&$\times$ & $\times$ & $\times$ & $\times$ & $\times$ & 
$\times$ & $-$ & $-$ & $-$ & $-$ & ${\tilde H}$ \\
het 5&$\times$ & $\times$ & $-$ & $-$ & $-$ & $-$ & 
$\times$ & $\times$ & $\times$ & $\times$ & $H$ \\
\end{tabular}}
\bigskip

\centerline{Diagram 1. The overlapping of two 5-branes.}
\bigskip\bigskip

    This solution can be resolved by introducing two sets of $SU(2)$
Yang-Mills instantons living in the Euclidean 4-spaces $dy_i^2$ and
$d\td y_i^2$ \cite{lima}.  Here we demonstrate that it can also be
resolved by a gravitational instanton together with matter $U(1)$
contributions.  Since the solution (\ref{hethet}) requires only that
$dy_i^2$ and $d\td y_i^2$ are Ricci-flat, we can replace each of them
with an Eguchi-Hanson metric, which we discuss in the appendix.  Since
the Eguchi-Hanson metric admits a self-dual (or anti-self-dual)
normalisable harmonic 2-form, we can turn on the matter $U(1)$ fields.
The solution now becomes
\bea
ds_{10}^2&=&(H\, {\wtd H})^{-\ft14}(-dt^2 +dw^2+H ds_{EH}^2+
{\wtd H} d{\td s}_{EH}^2) \,,\nn\\
{\rm e}^{-\phi}{\ast F}_\3&=&{\wtd H}\,dt\wedge 
dw\wedge \wtd \Omega_\4\wedge dH^{-1}+
H\, dt\wedge dw\wedge \Omega_\4\,\wedge d{\wtd H}^{-1}\,,\nn\\
\phi&=&\ft12 {\rm log}(H\,{\wtd H})\,,\qquad
F_\2^1 = m\, L_\2\,,\qquad
F_\2^2 =\td m\, \wtd L_\2\,,\label{reshethet}
\eea
where $\Omega_\4$ and $\wtd\Omega_\4$ are the volume forms for the
metric $ds_{EH}^2$ and $d\td s_{EH}^2$ respectively.  Here we have
turned on two $U(1)$ Cartan field strengths, labeled as $F_\2^1$ and
$F_\2^2$, living on $ds_{EH}^2$ and $d\td s_{EH}^2$ respectively.  Now
the Bianchi identity $dF_\3=\ft12 F_\2^1\wedge F_\2^1 + \ft12
F_\2^2\wedge F_\2^2$ implies that
\be
\square H = -\ft14\eta\, m^2\,L_\2^2\,,\qquad
\wtd {\square} \wtd H = -\ft14\td\eta\, \td m^2\, \wtd L_\2^2\,,
\ee
where $\eta^2=1=\td \eta^2$ are the orientation parameters of the
Eguchi-Hanson metrics (see the appendix). The equations of motion for
$F_\3$ and $F_\2^i$ are straightforwardly satisfied.  The dilaton
equation and the Einstein equation imply that $\eta=1=\td \eta$.  In
other words, although the equations of motion and Bianchi identity for
the form fields imply that the $L_\2^2$ and $\wtd L_\2^2$ source terms
can contribute both negatively or positively, depending on whether
they are self-dual or anti-self-dual, they are restricted to a
positive contribution due to the dilaton equation and Einstein
equation.  Thus, the Eguchi-Hanson instanton has to be such that its
normalisable harmonic 2-form is self-dual.  The generic solution for
$H$ and $\wtd H$ has a logarithmic divergent term, which vanishes for
appropriate integration constants \cite{clptrans}, giving
\be
H=1 + \fft{m^2}{2a^4\, r^2}\,,\qquad
\wtd H=1 + \fft{\td m^2}{2\td a^4\, \td r^2}\,,
\ee
where $a$ and $\td a$ are the sizes of the two Eguchi-Hanson
instantons.  Since the coordinates $r$ and $\td r$ run from $a$ and
$\td a$ to infinity respectively, it follows that the functions $H$
and $\wtd H$ are regular everywhere for non-vanishing $a$ and $\td a$.

       We have seen that the orientation of the Eguchi-Hanson metric
has a significant consequence in the resolution of the 5-branes.  The
positive orientation with $\eta=1$ leads to resolution whilst the
negative orientation does not solve all of the equations of motion.
We could have replaced $dy_i^2$ and $d{\td y}_i^2$ by a Taub-NUT
metric, in which case we would require the negative orientation with
$\eta=-1$ for resolution, since the normalisable harmonic 2-form has
to be self-dual.  The corresponding harmonic functions for the regular
solution become \cite{clptrans}
\be
H=1 + \fft{m^2}{4a\, (r+a)}\,,\qquad
\wtd H=1 + \fft{\td m^2}{4\td a\, (\td r +\td a)}\,.
\ee
For vanishing $m$ or $\td m$, the solution reduces to the
resolution of the heterotic 5-brane obtained in \cite{clptrans}.

       One can add a string along the common worldvolume of the above
configuration. The corresponding ``harmonic'' function $K$ satisfies
the equation \cite{cow}
\be
\wtd H\, \square K + H\, \wtd{\square}\, K=0\,,
\ee
which holds in our case as well.  The general solution of $K$ depends
on $H$ and $\wtd H$.  A natural special solution is $K=h\, \td h$,
where $h$ and $\td h$ are the harmonic functions of the two relative
four-dimensional transverse spaces respectively.  If the transverse
spaces are Euclidean, for the limit in which the gravitational
instanton sizes $a$ and $\td a$ vanish, then the three-component
solution has the near-horizon structure AdS$_3\times S^3\times
S^3\times E^1$ \cite{cow}.  The non-vanishing entropy of this
configuration disappears from the metric contribution once the
gravitational instanton is present. This suggests a phase transition
associated with the vanishing gravitational instanton; this is
analogous to the one associated with the Yang-Mills instantons for
overlapping heterotic 5-branes, discussed in \cite{lima}.  Note that
it is also possible to add a pp-wave component \cite{lima}.

    The present resolution of overlapping 5-branes incorporates
Yang-Mills fields, which are available only for heterotic string
theory.  On the other hand, unresolved overlapping 5-branes exist also
in type II theories.  While the resolution for the NS-NS and R-R
overlapping 5-branes of type II theories is also possible, there are
subtleties involved which will be discussed in section 4.

\section{Overlapping D4/D4 system}

     The D4-brane is a solution of type IIA supergravity, whose
bosonic Lagrangian is given by
\bea
{\cal L}_{\rm IIA}&=&R {\ast \oneone}-\ft12 {\ast d\phi} \wedge d\phi-
\ft12 {\rm e}^{-\phi}{\ast F}_\3\wedge F_\3-\ft12 {\rm e}^{\ft12\phi}
{\ast F}_\4 \wedge F_\4\nn\\
&&-\ft12{\rm e}^{\ft32\phi}{\ast F}_\2
\wedge F_\2-\ft12 dA_\3\wedge dA_\3\wedge A_\2\,,\label{IIA}
\eea
where
\be
F_\4=dA_\3-dA_\2\wedge A_\1\,,\qquad F_\3=dA_\2\,,\qquad
F_\2=dA_\1\,.\label{FIIA}
\ee

      The solution for a non-standard intersection of two D4-branes is
given by
\bea
ds_{10}^2 &=& -(H\, \wtd H)^{-\ft38}\, (dt^2 +
H\, dy_i^2 + \wtd H\, d\td y_i^2 +
H\, \td H\, dz^2)\,,\nn\\
{\rm e}^{\ft12\phi} {\ast F}_\4 &=&
\wtd H\, dt\wedge d^4\td y\wedge dH^{-1} +
H\, dt\wedge d^4 y\wedge d\wtd H^{-1}\,,\nn\\
\phi&=&-\ft14 {\rm log}(H\, \wtd H)\,,
\eea
where
\be
\square H=0\,,\qquad \wtd {\square} {\wtd H}=0\,,
\ee
and $\square$ ($\wtd {\square}$) is taken over the $y_i$ (${\td
y}_i$) directions.  This solution can be illustrated by the following
diagram:

\bigskip\bigskip
\centerline{
\begin{tabular}{c|ccccccccccc}
&$t$ & $y_1$ & $y_2$ & $y_3$ & $y_4$ & ${\tilde y}_1$ & ${\tilde y}_2$ &
${\tilde y}_3$ & ${\tilde y}_4$ & $z$ & \\ \hline
D4&$\times$ & $\times$ & $\times$ & $\times$ & $\times$ & $-$ & $-$ &
$-$ & $-$ & $-$ & $\wtd H_4$ \\
D4&$\times$ & $-$ & $-$ & $-$ & $-$ & $\times$ & $\times$ & $\times$ &
$\times$ & $-$ & ${H}_4$ \\
\end{tabular}}
\bigskip

\centerline{Diagram 2. The overlap of two D4-branes.}
\bigskip\bigskip

As with the heterotic 5-brane overlap, we replace the Euclidean
4-spaces $dy_i^2$ and $d\td y_i^2$ by Eguchi-Hanson metrics
$ds_{EH}^2$ and $d\td s_{EH}^2$ respectively.  By making use of the
Bianchi identity, $dF_\4=F_\3 \wedge F_\2$, we can consider the
following ansatz
\bea
ds_{10}^2 &=& -(H\, \wtd H)^{-\ft38}\, (dt^2 +
H ds_{EH}^2 + \wtd H d\td s_{EH}^2 +
H \td H dz^2)\,,\nn\\
{\rm e}^{\ft12\phi} {\ast F}_\4 &=&
\wtd H\, dt\wedge \wtd \Omega_\4\wedge dH^{-1} +
H\, dt\wedge \Omega_\4\wedge d\wtd H^{-1}\,,\\
\phi&=&-\ft14 {\rm log}(H\, \wtd H)\,,\quad
F_\3=(m\, L_\2 - \td m\, \wtd L_\2)\wedge dz\,,\quad
F_\2=m\, L_\2 + \td m\, \wtd L_\2\,.\nn
\eea
$L_\2$ is a normalisable self-dual harmonic 2-form in $ds_{EH}^2$,
which hence has positive orientation $\eta=1$, and $\wtd L_\2$ is a
normalisable anti-self-dual harmonic 2-form in $d\td s_{EH}^2$, which
hence has negative orientation $\eta=-1$.  Since we have
\be
F_\4 = ({\td \ast_4 d\wtd H} + {\ast_4 dH})\wedge dz\,,
\ee
where $\td\ast_4$ and $\ast_4$ are the Hodge duals in $d{\td
s}_{EH}^2$ and $d s_{EH}^2$ respectively, the Bianchi identity for
$F_\4$ implies that
\be
\square H = -\ft12 m^2\,L_\2^2\,,\qquad
\wtd {\square} \wtd H = -\ft12\td m^2\, \wtd L_\2^2\,.\label{d4d4h}
\ee
It should be mentioned that the different choice of sign in $F_\3$ and
$F_\2$ ensures that the cross term in $F_\3\wedge F_\2$ cancels out.
Due to the minus sign of $\wtd L_\2$ in $F_\3$, it contributes to
$\wtd H$ in the same way that $L_\2$ contributes to $H$, even though
$\wtd L_\2$ is anti-self-dual and $L_\2$ is self-dual.

     To check the rest of the equations of motion, it is useful to
present the following:
\bea 
{\rm e}^{-\phi} {\ast F_\3} &=& H^{-1}\,\wtd H\, dt\wedge \wtd
\Omega_\4\wedge L_\2 + H\, \wtd H^{-1}\, dt\wedge \Omega_\4\wedge \wtd
L_\2\,,\nn\\
{\rm e}^{\ft32\phi} {\ast F_\2} &=& H^{-1}\,\wtd H\,
dt\wedge \wtd \Omega_\4\wedge dz\wedge L_\2 - H\, \wtd H^{-1}\,
dt\wedge \Omega_\4\wedge dz\wedge \wtd L_\2\,,\nn\\
{\ast d\phi} &=&
\ft14 H^{-1}\, \wtd H\, dt\wedge \wtd \Omega_\4 \wedge {\ast_4
dH}\wedge dz + \ft14 H\, \wtd H^{-1}\, dt\wedge \Omega_\4\wedge
{\td\ast_4 d\wtd H}\wedge dz\,.\label{d4d4res}
\eea 
We verify that all of the equations of motion for the form fields and
the dilaton are satisfied.  We did not verify the Einstein equation
due to its complexity for these cases, although past experience leads
us to believe that it is satisfied.  The regular solution for
(\ref{d4d4h}) is given by
\be
H=1 + \fft{m^2}{a^4\, r^2}\,,\qquad
\wtd H=1 + \fft{\td m^2}{\td a^4\, \td r^2}\,.
\ee
Each of the Eguchi-Hanson metrics can also be replaced by a Taub-NUT
metric with proper orientation such that it has the required
normalisable self-dual or anti-self-dual harmonic 2-forms.  The
resulting regular solutions for $H$ and $\td H$ are given by
\be
H=1 + \fft{m^2}{2a\, (r+a)}\,,\qquad
\wtd H=1 + \fft{\td m^2}{2\td a\, (\td r +\td a)}\,.
\ee

       If $m$ or $\td m$ vanishes, then the solution becomes that of a
single resolved D4-brane with the transverse space $ds_5^2 = ds_4^2 +
dz^2$, where $ds_4^2$ is a Ricci-flat manifold that admits
normalisable self-dual or anti-self-dual harmonic 2-forms $L_\2$.  In
this case, we have $F_\2=m\, L_\2$ and $F_\3=m\, {\ast_5 L_\2}$.
Thus, we see that $F_\3\wedge F_\2$ always contributes positively to
the Bianchi identity for the $F_\4$.  Both orientations of
Eguchi-Hanson instanton or Taub-NUT can be used to resolve the
D4-brane.  This is very different from the case of the heterotic
5-brane, in which only one orientation can be used to resolve the
brane.  In order to cancel out the cross term in $F_\3\wedge F_\2$, we
find that $ds_4^2$ and $d\td s_4^2$ must have opposite orientations
such that they admit normalisable self-dual and anti-self-dual
harmonic 2-forms respectively.

\section{Overlapping type II 5-branes}

        Overlaps of type II NS-NS or R-R 5-branes share the same
metric structure as that of overlapping heterotic 5-branes,
illustrated in Diagram 1.  However, the resolution of the heterotic
5-brane is rather unique, since it makes use of multiple matter
Yang-Mills fields which are absent in the type II theories.  For this
reason, the resolution of the type II 5-brane was previously unknown.
A regular solution of the 5-brane wrapped around $S^2$ was obtained in
\cite{mn} by lifting the four-dimensional $SU(2)$ gauged black hole
\cite{cv}.  This solution can apply for both type II and heterotic
5-branes.  In this section, we obtain a resolved type II 5-brane
overlap by performing T-duality on the D4/D4 system of the previous
section.  By turning off one component, we obtain a resolved 5-brane
wrapped on $S^1$.

       The resolved D4/D4 solution (\ref{d4d4res}) has a $U(1)$
isometry in the $z$ direction.  By performing T-duality on this
direction, we obtain the overlap of two (R-R) D5-branes of type IIB
supergravity:
\bea
ds_{10}^2&=&(H\, {\wtd H})^{-\ft14}(-dt^2 +(dz+\cA_\1)^2+H ds_{EH}^2+
{\wtd H} d{\td s}_{EH}^2) \,,\nn\\
F_\3^{\rm RR} &=& {\ast_4 dH} + {\td \ast_4  d\wtd H}
- (m\, L_\2 + \td m\, \wtd L_\2)\wedge (dz+ \cA_\1)\,,\nn\\
\phi&=&-\ft12 {\rm log}(H\,{\wtd H})\,,\qquad d\cA_\1=
m\, L_\2 - \td m\, \wtd L_\2\,.\label{2a55}
\eea
The resolved NS-NS overlapping 5-brane can be easily obtained by
performing the S-duality of the type IIB theory, with the $F_\3^{\rm
RR}$ replaced by $F_\3^{\rm NS}$ and the sign of the dilaton changed.
Since the solution involves only the metric, dilaton and the 3-form
field strength, it is also valid for resolving the 5-branes in type
IIA or heterotic strings (without a Yang-Mills source).  It is
interesting to note that there is a twist along the direction $z$.
The topology of a spatial slice of the solution can be viewed as a
$U(1)$ fibration of the product space of two Eguchi-Hanson instantons
with opposite orientations.  The solution describes that a regular
effective string, as common worldvolume of two 5-branes, wraps on the
fibre circle $z$.  Note that unlike the previous examples this
resolution makes use of only the interaction between the gravity and
the 3-form field strength, instead of the interactions associated with
the Bianchi identities or the equations of motion of the form fields.

        If $m$ or $\td m$ vanishes, then the solution becomes a
resolved single 5-brane. Without the loss of generality, we set $\td
m=0$.  The metric solution becomes
\bea
ds_{10}^2&=& H^{-\ft14} (-dt^2 + dx_1^2 + \cdots + dx_4^2 +
(dz + \cA_\1)^2) + H^{\ft34}\, ds_{EH}^2\,,\label{5braneres}
\eea
with $d\cA_\1=m\, L_\2$.  The expressions for the 3-form field
strength and the dilaton can be easily read off from (\ref{2a55}).
The solution describes 5-branes wrapped around $S^1$, which is fibred
over the transverse space $ds_{EH}^2$.  It is regular everywhere,
providing a well-behaved supergravity dual of a certain $D=5$ super
Yang-Mills theory.

      It is important to note that the resolved NS-NS 5-brane as well
as the resolved overlap of two NS-NS 5-branes are invariant under a
T-duality transformation along the $z$ direction.  This is rather
different from the usual situation where T-duality would untwist the
fibration \cite{dlp1}.  (Analogous phenomenum was observed in the
NS-NS dyonic string \cite{dlp2}.)  Invariance under T-duality is a
property of NS-NS 5-branes. Since T-duality is a symmetry at all
orders of perturbative string theory, we would expect that the
invariance should hold for the resolved solution, and that is indeed
the case.

      If instead the resolved 5-brane (\ref{5braneres}) carries the
type IIB R-R $F_\3^{\rm RR}$ charge, we can perform T-duality on $x_4$
and obtain the resolved D4-brane of type IIA
\bea
ds_{10}^2 &=& H^{-\ft38}\, (-dt^2 + dx_1^2 + dx_2^2 + dx_3^2 +
(dz+\cA_\1)^2) + H^{\ft58}\, (ds_{EH}^2 + d\td z^2)\,,\\
F_\4&=& ({\ast_4dH} - m\, L_\2\wedge (dz+\cA_\1))\wedge d\td
z\,,\qquad \phi=-\ft14\log(H)\,,\qquad d\cA_\1=m\, L_\2\,.\nn
\eea
Since this solution describes D4-brane wrapped on $S^1$, which is
fibred over the tranverse space $ds_{EH}^2 + d\td z^2$, it has
an effective four-dimensional world-volume.  Thus we find a new
well-behaved supergravity solution dual to a certain ${\cal N}=2$,
$D=4$ supersymmetric Yang-Mills theory.

\section{Conclusions}

      The resolution of singularities in supergravity BPS brane
solutions provides a convenient way of extending the validity of these
solutions.  In this letter, we have considered the resolutions of two
overlapping heterotic 5-branes, type II 5-branes or D4-branes.  The
relative transverse spaces in these overlapping solutions are all
four-dimensional and hence can be replaced by either Eguchi-Hanson or
Taub-NUT spaces, both of which admit normalisable self-dual or
anti-self-dual harmonic 2-forms, depending on the orientation.  Terms
corresponding to interactions between form fields modify Bianchi
identities and equations of motion.  It follows that these
normalisable harmonic 2-forms provide regular sources for the branes.

      When each of the two relative transverse Euclidean 4-spaces is
replaced by the Eguchi-Hanson or the Taub-NUT instanton, half of the
supersymmetry is broken.  Introducing a brane configuration will not
break the supersymmetry any further.  Thus our resolved overlapping
brane solutions preserve $\ft14$ of the supersymmetry.

      We have also obtained resolved 5-branes and D4-branes wrapped on
$S^1$, which is fibred over the transverse Eguchi-Hanson or Taub-NUT
spaces.  This provides a regular supergravity dual to a certain $D=5$
and $D=4$ super Yang-Mills theory.

      The resolved solution are regular everywhere in the spacetime,
with a stable dilaton and hence a stable string coupling constant.
Furthermore, unlike a typical brane solution that requires a brane
source term that is beyond supergravity, the resolved ones are
complete purely within supergravity.  Not all the BPS branes can be
resolved at the level of supergravity.  It is interesting to find
those that can be resolved and study the special role they play in
string and M-theory.

\section*{Appendix}

\appendix

\section{Normalisable harmonic forms in $D=4$}

\subsection{Eguchi-Hanson Instanton}

   Let us consider the Eguchi-Hanson solution \cite{eh}
\bea
ds_4^2&=& W^{-1}\, dr^2 + \ft14 r^2\, W\, (d\psi + \eta\, \cos\theta\,
d\phi)^2 + \ft14 r^2\, (d\theta^2 + \sin^2\theta\, d\phi^2)\,,\nn\\
W&=& 1 -\fft{a^4}{r^4}\,,\label{eh}
\eea
The radial coordinate $r$ lies in the range $a\le r\le \infty$, and
$\psi$ has period $2\pi$ \cite{bgpp} to ensure regularity at $r=a$.
Thus, the metric is asymptotically locally Euclidean (ALE), with the
periodicity condition on $\psi$ implying that constant $r$ surfaces
are $RP^3=S^3/Z_2$ \cite{bgpp}.  The coordinate $\psi$ can have two
orientations of the fibration, corresponding to $\eta=\pm 1$.  The
metric is K\"ahler with the vielbein basis
\be
e^0=W^{-\ft12}\, dr\,,\quad
e^1=\ft12r\, d\theta\,,\quad
e^2=\ft12 r\, \sin\theta\, d\phi\,,\quad
e^3=\ft12 r\, W^{\ft12}\, (d\psi+\eta\,\cos\theta\, d\phi)\,,
\ee
and with a K\"ahler form given by
\be
J=e^0\wedge e^3 - \eta\, e^1\wedge e^2\,.
\ee
It is anti-self-dual for $\eta=1$ and self-dual for $\eta=-1$.  The
metric is also admits a harmonic 2-form, given by \cite{clptrans}
\be
L_\2 = \fft{2}{r^4} (e^0\wedge e^3 + \eta\, e^1\wedge e^2)\,,
\ee
which is self-dual for $\eta=1$ and anti-self-dual for $\eta=-1$.  The
square of $L_\2$ is given by $L_\2^2=16/r^8$, and hence this 2-form is
normalisable.   Note that we have
\be L_\2\wedge L_\2 = \ft12 \eta\, L_\2^2\, \Omega_\4\,, 
\ee 
where $\Omega_\4$ is the volume form of the metric (\ref{eh}).  It
follows that the sign of $\eta$ is not merely a choice of convention
but rather it has a non-trivial physical consequence.

\subsection{Taub-NUT instanton}

      The metric of the Taub-NUT instanton is given by \cite{hawking}
\be
ds_4^2 = \Big(\fft{r+a}{r-a}\Big)\, dr^2 + 4a^2\,
\Big(\fft{r-a}{r+a}\Big)
(d\psi + \eta\,\cos\theta\, d\phi)^2 + (r^2-a^2)\, (d\theta^2 +
\sin^2\theta\, d\phi^2)\,,
\ee
where the radial coordinate runs from $r=a$ to $r=\infty$, and $\psi$
has period $4\pi$ to guarantee that the solution is regular at $r=a$.
Note that $\eta=\pm 1$ are the two orientation choices.  The Taub-NUT
manifold has the topology $\R^4$ but, even though the metric at large
$r$ is asymptotically flat, it approaches the cylinder $\R^3\times
S^1$ rather than Euclidean space.  The orthonormal frame is given by
\bea
&&e^0= \Big(\fft{r-a}{r+a}\Big)^{-1/2}\, dr\,,\quad e^1=
(r^2-a^2)^{1/2}\, d\theta\,, \qquad e^2 = (r^2-a^2)^{1/2}\,
\sin\theta\, d\phi\,,\nn\\
&&e^3 = 2a\, \Big(\fft{r-a}{r+a}\Big)^{1/2}\,
(d\psi + \eta\, \cos\theta\, d\phi)\,.
\eea
There is one normalisable harmonic 2-form, given by \cite{clptrans}
\be
L_\2=\fft{1}{(r+a)^2}(e^0\wedge e^3 - \eta\, e^1\wedge e^2)\,,
\ee
which is self-dual for $\eta=-1$ and anti-self-dual for $\eta=+1$.
It is normalisable since
\be
(L_\2)^2 = \fft{4}{(r+a)^4}\,.
\ee

\section*{Acknowledgments}

We would like to thank Xavier Bekaert, Nicolas Boulanger, Mirjam
Cveti\v{c} and Chris Pope for useful conversations.

\vfil\eject

\end{document}